\documentclass[12pt,letterpaper]{article}
\usepackage[top=0.85in,left=2.75in,footskip=0.75in,marginparwidth=2in]{geometry}
\usepackage{amsmath,amsthm,amssymb,amsfonts,enumerate,ae,hyperref,graphicx,cite,dsfont,bbm,bbold,setspace,subfigure,pgf}

\usepackage[utf8]{inputenc}

\usepackage{cite}

\usepackage{nameref,hyperref}

\usepackage{microtype}
\DisableLigatures[f]{encoding = *, family = * }

\raggedright
\usepackage{changepage}

\usepackage[aboveskip=1pt,labelfont=bf,labelsep=period,singlelinecheck=off]{caption}

\makeatletter
\renewcommand{\@biblabel}[1]{\quad#1.}
\makeatother

\usepackage{lastpage,fancyhdr,graphicx}
\usepackage{epstopdf}
\fancyheadoffset[L]{2.25in}
\fancyfootoffset[L]{2.25in}

\usepackage{color}

\definecolor{Gray}{gray}{.25}

\usepackage{graphicx}

\usepackage{sidecap}

\usepackage{wrapfig}
\usepackage[pscoord]{eso-pic}
\usepackage[fulladjust]{marginnote}
\reversemarginpar

\begin{document}
\vspace*{0.35in}

\begin{flushleft}
{\Large
\textbf\newline{Altered Topological Structure of the Brain White Matter in Maltreated Children through Topological Data Analysis}
}
\newline
\\
Moo K. Chung\textsuperscript{1,*},
Tahmineh Azizi\textsuperscript{1},
Jamie L. Hanson\textsuperscript{2},
Andrew L. Alexander\textsuperscript{4},
Richard J. Davidson\textsuperscript{5},
Seth D. Pollak\textsuperscript{5}
\\
\bigskip
\bf{1} Department of Biostatistics and Medical Informatics, University of Wisconsin-Madison, USA
\\
\bf{2} Department of Psychology,  University of Pittsburgh, USA
\\
\bf{4} Department of Medical Physics, University of Wisconsin-Madison, USA
\\
\bf{5} Department of Psychology, University of Wisconsin-Madison, USA
\bigskip

*Correspondence: Moo K. Chung, 
Email: {\tt mkchung@wisc.edu}. 
\end{flushleft}

\section*{Abstract}

Childhood maltreatment may adversely affect brain development and consequently influence behavioral, emotional, and psychological patterns during adulthood. In this study, we propose an analytical pipeline for modeling the altered topological structure of brain white matter in maltreated and typically developing children. We perform topological data analysis (TDA) to assess the alteration in the global topology of the brain white-matter structural covariance network  among children. We use persistent homology, an algebraic technique in TDA, to analyze topological features in the brain covariance networks constructed from structural magnetic resonance imaging (MRI) and diffusion tensor imaging (DTI). We develop a novel framework for statistical inference based on the Wasserstein distance to assess the significance of the observed topological differences. Using these methods in comparing maltreated children to a typically developing control group, we find that maltreatment may increase homogeneity in white matter structures and thus induce higher correlations in the structural covariance; this is reflected in the topological profile. 
Our findings strongly suggest that TDA can be a valuable framework to model altered topological structures of the brain. The MATLAB codes and processed data used in this study can be found at 
\url{https://github.com/laplcebeltrami/maltreated}.

\section{Introduction}

Child maltreatment can have severe life-long mental, emotional, physical, and sexual health outcomes \cite{WHO2022child}. These serious long-term consequences are notable given that the U.S. Department of Health and Human Services estimates over 680,000 children suffer different forms of maltreatment, such as child abuse or neglect every year. Many of the adverse impacts likely emerge through changes in neurobiology, such as reduced brain volumes and altered brain connectivity \cite{herringa2013childhood}. Indeed, a growing body of scientific research has found altered brain functioning in those who have suffered early childhood abuse and neglect \cite{hanson2010early,mccrory2010research,shonkoff2012lifelong,wilson2011traumatic}. Multiple studies have shown that maltreatment in childhood can lead to a decrease in the volume of the corpus callosum, the largest white matter structure in the brain, which is critical for interhemispheric communication \cite{mccrory2010research,wilson2011traumatic}. Similarly, neglected children tend to have smaller prefrontal cortex volumes, which play a role in regulating behavior, emotion, and cognition \cite{NSCD2010child, NSCD2010child2}. These neurological changes, especially those in brain connectivity, may profoundly influence children's emotional, social, and behavioral functioning \cite{hostinar2012associations,NSCD2010child}.

Both structural MRI and diffusion MRI facilitate studies on the impact of abuse and neglect on brain development during childhood \cite{pollak2008mechanisms,loman2013effect,hanson2012structural,jackowski2009neurostructural}. Tensor-based morphometry (TBM) serves as a powerful tool to quantify the variations in neuroanatomical structures by analyzing the spatial derivatives of deformation fields. These fields are obtained via nonlinear image registration techniques that warp individual structural MRI scans to a common template \cite{thompson1998growth,chung2001unified}. The Jacobian determinant, derived from this warping process, measures the volumetric changes in brain tissue at the voxel level \cite{davatzikos1996computerized,dubb2003characterization,machado1998atlas}. For each voxel, a linear model is set up to use tensor maps, such as the Jacobian determinant, as a response variable for obtaining voxel-level statistics. Although univariate TBM has been widely utilized \cite{chung2001unified,thompson1998anatomically}, its limitations emerge when hypothesis testing extends to multiple anatomical brain regions; it may not adequately capture the inter-relationships between volume changes in different voxels. This gap underscores the need for a network analysis approach to model the Jacobian determinant, linking variations in one region to another through structural covariance \cite{cao1999geometry,he2008structural,he2007small,lerch2006mapping,rao2008hierarchical,worsley2005connectivity,worsley2005comparing}.

Keith Worsley laid the foundation for modeling structural covariance using cortical thickness obtained from T1-MRI in 2005 \cite{lerch.2006,worsley.2005.neural,worsley2005connectivity,worsley2005comparing}. Worsley's contributions were instrumental in framing the concept of structural covariance as the statistical association between morphological characteristics of different brain regions. His work inspired a wealth of research that employed statistical models to quantify these associations. After Worsley's initial contributions, the field saw significant developments  with greater sophistication \cite{lerch2006mapping,he2007small,he2008structural}. In early 2010's, studies began to explore the application of structural covariance in various neurological and psychiatric conditions, such as Alzheimer's disease (AD), schizophrenia, and developmental disorders  including fragile X syndrome \cite{rao2008hierarchical,cao1999geometry,saggar.2015}. \cite{dupre.2017} used the gray matter probability map obtained from the SPM package in modeling the lifespan of structural covariance networks in the normal population. These studies often employed machine learning and network theory to create more complex models that could capture the intricate relationships between different brain regions. Most recently, the notion of structural covariance has been integrated into multimodal imaging studies, which combine different types of neuroimaging data to provide a more comprehensive view of brain structure and function \cite{chung.2015.TMI,machado1998atlas,davatzikos1996computerized}.

Graph theory based methods have been frequently used to uncover the topological properties of brain networks including the investigation of topological alterations in white matter for neuromyelitis optica \cite{liu2012altered}, exploring abnormal topological organization in the structure of cortical networks in AD \cite{lo2010diffusion}, alterations in the topological properties of the anatomical network in early blindness \cite{shu2009altered},  abnormal topological changes during AD progression  \cite{kuang2020white,daianu2015rich,qiu2016disrupted}. Graph theory also has been used to measure and evaluate the integration and segregation of the brain network \cite{rubinov2010complex,kuang2020white}. In the standard graph theory based brain network analysis,  graph features such as node degrees and clustering coefficients are obtained after thresholding connectivity matrices \cite{van2010comparing,chung2017integrative,chung2019brain}. Depending on the choice of these thresholds, the final statistical results can be drastically different \cite{lee.2012.tmi,chung2013persistent,chung2015persistent}. Thus, there is a practical need to develop a {\em multiscale} network analysis framework that provides  a consistent  result and interpretation regardless of the choice of thresholding. Persistent homology offers one possible solution to the multiscale problem \cite{chung2013persistent,carlsson2008persistent,chung2009topological,edelsbrunner2000topological,ghrist2008barcodes,lee2011computing,lee.2012.tmi,singh2008topological}.

Persistent homology has gained popularity for its capability to analyze high dimensional feature spaces without model assumptions \cite{edelsbrunner2000topological,ghrist2008barcodes,chung2009topological,lee2011computing}. Instead of studying networks at a fixed scale, persistent homology summarizes the changes of topological features over different scales and finds the most persistent topological features that are robust to  perturbations \cite{chung.2019.NN}. This robust performance under different scales is needed for network models that are parameter and scale dependent. In persistent homology, instead of building networks at one fixed parameter that may not be optimal \cite{ghrist2008barcodes,edelsbrunner2008persistent}, we analyze the collection of networks over every possible thresholds \cite{lee2011computing,lee.2012.tmi}. It has been shown that the persistent homology approach can be effectively used to overcome the problem related to the arbitrariness of thresholding  \cite{lee2019clinical}. Persistent homology can detect subtle topological differences between networks while existing statistical models might fail to differentiate the differences  \cite{zhu2014matrix,qiu2015manifold,solo2018connectivity}. In \cite{liu2021neurophysiological}, persistent homology has been applied to characterize the neuropsychological properties of the brain. In \cite{xing2022spatiotemporal}, persistent homology has been used to study the evolution of a spatiotemporal brain network of Alzheimer’s disease (AD). They have also proposed that persistent homology can be considered as a framework to assess the neurophysiological properties of image quality. Topological data analysis (TDA) has been applied to brain networks to classify altered brain states \cite{caputi2021promises}. TDA also has been used to extract the topology of brain connectomes in attention deficit hyperactivity disorder (ADHD) \cite{gracia2020topological}. TDA also found applications in EEG signal analysis \cite{piangerelli2018topological,wang2019statistical,khalid2014tracing}.

Various topological feasters and embedding have been developed. The persistence diagram (PD) serves as an indicator, displaying the birth and death times of holes or cycles as the scale changes. Important topological invariants, known as Betti numbers, count the number of holes in networks and can be used to visualize and quantify underlying topology. Betti curves, which plot these Betti numbers over changing scales, have been employed to detect abnormal functional brain networks in the study of Alzheimer's Disease (AD) progression \cite{kuang2020white}. Furthermore, a variety of quantitative persistent homology features exist, such as persistence landscapes (PL) \cite{bubenik2017persistence}, persistent entropy (PE) \cite{rucco2016characterisation}, and persistence images (PI) \cite{adams2017persistence}. These features have been utilized to analyze and compare brain networks across different patients \cite{caputi2021promises}. Mapper is another commonly used TDA technique, particularly useful for simplifying high-dimensional data into network representations by providing insights into the clustering and connectedness of data points in a feature space \cite{patania.2019,saggar.2018}. Mapper can be effective in capturing the network modularity and revealing the hierarchical organization of functional brain connectivity \cite{patania.2019}.  \cite{saggar.2018} used Mapper to construct the low-dimensional representations of temporally changing task fMRI brain networks.  \cite{petri.2014} introduces the clique filtration in building homological scaffolds  that serve as the backbone for understanding the topological organization of  fMRI brain networks. These tools are particularly useful in capturing the intricate higher-order topological features, such as loops and voids, that are often not readily accessible in existing methods.

 In this study, we use TDA to investigate alterations in the white matter structures of children who have experienced maltreatment. Utilizing both T1-MRI and DTI scans, we focus on the structural covariance of the brain's white matter. Techniques from persistent homology are employed to characterize these changes, specifically using the Jacobian determinant from tensor-based morphometry (TBM) and fractional anisotropy (FA) values from DTI. Unlike univariate-TBM, persistent homology enables us to examine more intricate network hypotheses, capturing subtle variations across voxels. We quantify these topological properties using Betti curves and apply the Wasserstein distance to differentiate between maltreated and control groups. This methodology allows us to robustly characterize topological structures at multiple scales. Our results reveal that maltreated children exhibit significant alterations in white matter topology compared to controls, including a lower number of connected components, suggesting less heterogeneous white matter structures.

\section{Methods}

\begin{figure}[t]
	\centering
	\includegraphics[width=1\linewidth]{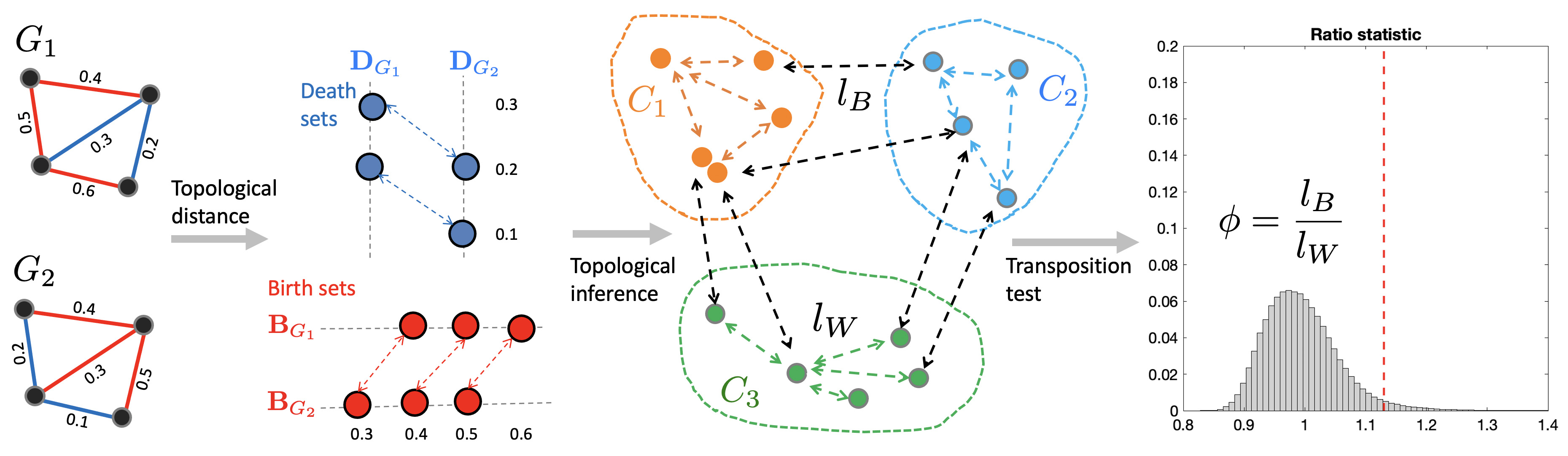}
	\caption{Proposed topological inference pipeline for analyzing structural covariance networks. Given two weighted graphs \( G_1, G_2 \), we first perform the birth-death decomposition and partition the edges into sorted birth and death sets (section \ref{sec:BDD}). 
The 0D topological distance between birth values quantifies discrepancies in connected components (section \ref{sec:distance}). 
The 1D topological distance between death values quantifies discrepancies in cycles. Topological inference is based on the ratio of between-group distance $l_B$ to within-group distance $l_W$ (section \ref{sec:inference}). Statistical significance on the ratio $\phi= l_B /l_W$  is assessed using the transposition test, a scalable online permutation test.}
	\label{fig:method1}
\end{figure}

Figure \ref{fig:method1} displays the overall pipeline for group level network analysis. Even though the method is applied to structural covariance networks, it works for any type of networks as long as the networks are represented as weighted graphs.

\subsection{Birth and death decomposition}
\label{sec:BDD}

In this study, we represent a brain network as weighted graph \( G = (V, w) \), where \( V = \{1, 2, \ldots, q\} \) is the node set and \( w = (w_{ij}) \) denotes edge weights, yielding \( r = (q^2 - q)/2 \) total edges \cite{lee.2012.tmi,petri.2014}. The weighted graph can be treated as simplicial complexes \cite{edelsbrunner2022computational,zomorodian2005topology}. One commonly used simplicial complex is the Rips complex \( \mathcal{R}_{\epsilon} \), defined as consisting of \( k \)-simplices formed by \( k+1 \) nodes within distance \( \epsilon \) \cite{ghrist2008barcodes}. For a graph with \( q \) nodes, the Rips complex can contain simplices up to dimension \( q - 1 \).
Then the hierarchical nesting structure called the {\em Rips filtration} is induced by the Rips complex:
$$
	R_{\epsilon_{0}} \subset R_{\epsilon_{1}}\subset R_{\epsilon_{2}} \subset \dots
$$
where $ 0=\epsilon_{0}<\epsilon_{1}<\epsilon_{2} <\dots$ are called the filtration values. 
When the number of nodes becomes large, the Rips complex becomes very dense and often causes serious computational bottlenecks in  computationally demanding tasks  such as the permutation test. For this reason, we propose to use the graph filtration, a special case of Rips filtration restricted to 1-skeleton \cite{lee.2012.tmi,lee2011computing}.

\begin{figure}[h]
	\centering
	\includegraphics[width=1\linewidth]{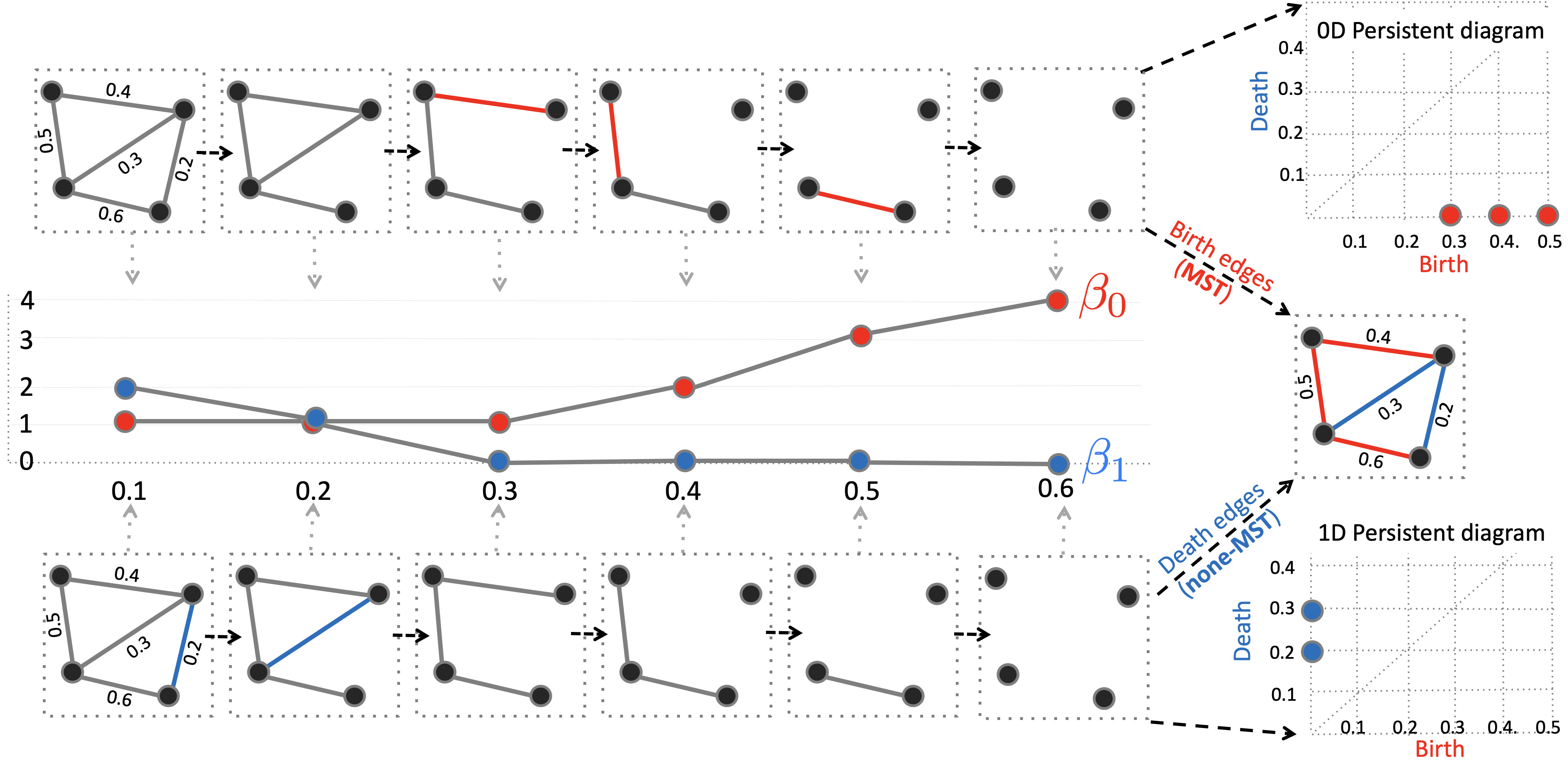}
	\caption{Illustration of a graph filtration with corresponding birth-death decomposition. During the graph filtration, edges are removed one at a time, starting from the smallest edge weight to the largest. Each edge removal either creates a new connected component (highlighted in red) or eliminates a cycle (highlighted in blue). The parameter \(\beta_0\), which counts the number of connected components, is monotonically non-decreasing, while \(\beta_1\), which counts the number of cycles. Thus, the edges can be decomposed into birth and death sets: the birth set corresponds to the maximum spanning tree (MST), and the death set comprises non-MST edges. The birth set forms the 0D persistence diagram, while the death set forms the 1D persistence diagram.}
	\label{fig:GF}
\end{figure}

Define the binary graph $G_{\epsilon}=(V,w_{\epsilon}) $ with binary edge weights $ w_{\epsilon}=(w_{\epsilon,ij}) $ such that
$$
  w_{\epsilon,ij} =
\begin{cases}
        1 & \text{ for }  w_{ij}>\epsilon ,   \\
   0 & \text{ otherwise.}
  \end{cases}  
$$
The binary matrix $ w_{\epsilon} $ is  the adjacency matrix of $ G_{\epsilon}$ and defines a simplicial complex only using  0-simplices (nodes) and 1-simplices (edges) \cite{lee.2012.tmi}. We then obtain the graph filtration of $G$ as a sequence of nested multiscale binary graphs:
$$
	G_{\epsilon_{0}} \supset G_{\epsilon_{1}}\supset \dots \supset G_{\epsilon_{k}}
$$
with filtration values $ \epsilon_{0}<\epsilon_{1}<\epsilon_{2} <\dots <\epsilon_{k} $ \cite{lee2011computing}. Figure \ref{fig:GF} displays an example of  graph filtration with four nodes.

Change in the filtration values \( \epsilon \) may cause the appearance or disappearance of connected components or loops \cite{chung.2019.NN}. In a simplicial complex, the number of connected components is the Betti-0 number \( \beta_{0} \), and the number of independent cycles (or loops) is the Betti-1 number \( \beta_{1} \). In graph filtrations, \( \beta_{0} \) increases while \( \beta_{1} \) decreases over filtrations (Figure \ref{fig:GF}) \cite{chung.2019.NN}. During the graph filtration, a connected component that is born never dies; thus, the death time is infinity. Consequently, we ignore the death values of connected components and characterize them by a set of increasing birth values \( {\bf B}_G \): 
$$
    {\bf B}_G:\epsilon_{b_{1}} < \dots < \epsilon_{b_{m_{0}}}.
$$
On the other hand, loops are always present in complete graphs, so the birth values of cycles are considered as \( -\infty \) and are ignored.  The loops are then completely characterized by a set of increasing death values \( {\bf D}_G \): 
$$
    {\bf D}_G:\epsilon_{d_{1}} < \dots < \epsilon_{d_{m_{1}}}.
$$
Thus, we can decompose edge weights \( w=(w_{ij}) \) uniquely into either the birth set \( {\bf B}_G \) or death set \( {\bf D}_G \) through the {\em birth-death decomposition} \cite{song.2023}:
$$
        w={\bf B}_G \cup {\bf D}_G, \quad {\bf B}_G \cap {\bf D}_G \neq \emptyset ,
$$
where \( {\bf B}_G=\{\epsilon_{b_{1}},\epsilon_{b_{2}},\dots , \epsilon_{b_{m_{0}}}\} \) and \( {\bf D}_G=\{\epsilon_{d_{1}},\epsilon_{d_{2}},\dots , \epsilon_{d_{m_{1}}}\} \) with \( m_{0}=q-1 \) and \( m_{1}=(q-1)(q-2)/2 \). The birth set \( {\bf B}_G \) is equivalent to the maximum spanning tree (MST) of \( G \) and forms the persistent diagram for 0D homology (connected components). On the other hand, the death set \( {\bf D}_G \) consists of edges that do not belong to the MST and forms the persistent diagram for 1D homology (cycles).  We compute the Betti-0 curves using Kruskal's algorithm, which works by identifying the minimum spanning tree to construct Betti-0 curves \cite{lee.2012.tmi}. Betti-1 curves are then identified through the Euler characteristic \cite{chung.2019.NN,chung2019statistical}. The computation can be done in \( \mathcal{O}( q^2 \log q) \) runtime. The computation is done through MATLAB function call
{\tt [Wb Wd] = WS\_decompose(W)}, which inputs the connectivity matrix $W$ and outputs the birth set {\tt Wb} and the death set {\tt Wd}.

\subsection{Wasserstein distances between networks}
\label{Dist}
\label{sec:distance}

The topological distance between persistence diagrams is often measured using the 2-Wasserstein distance. For graph filtrations, the persistence diagrams consist of 1D sorted birth or death values. Thus, the Wasserstein distance can be computed through order statistics on edge weights \cite{das.2023,song.2023}.

\begin{figure}[t]
\centering
\includegraphics[width=1\linewidth]{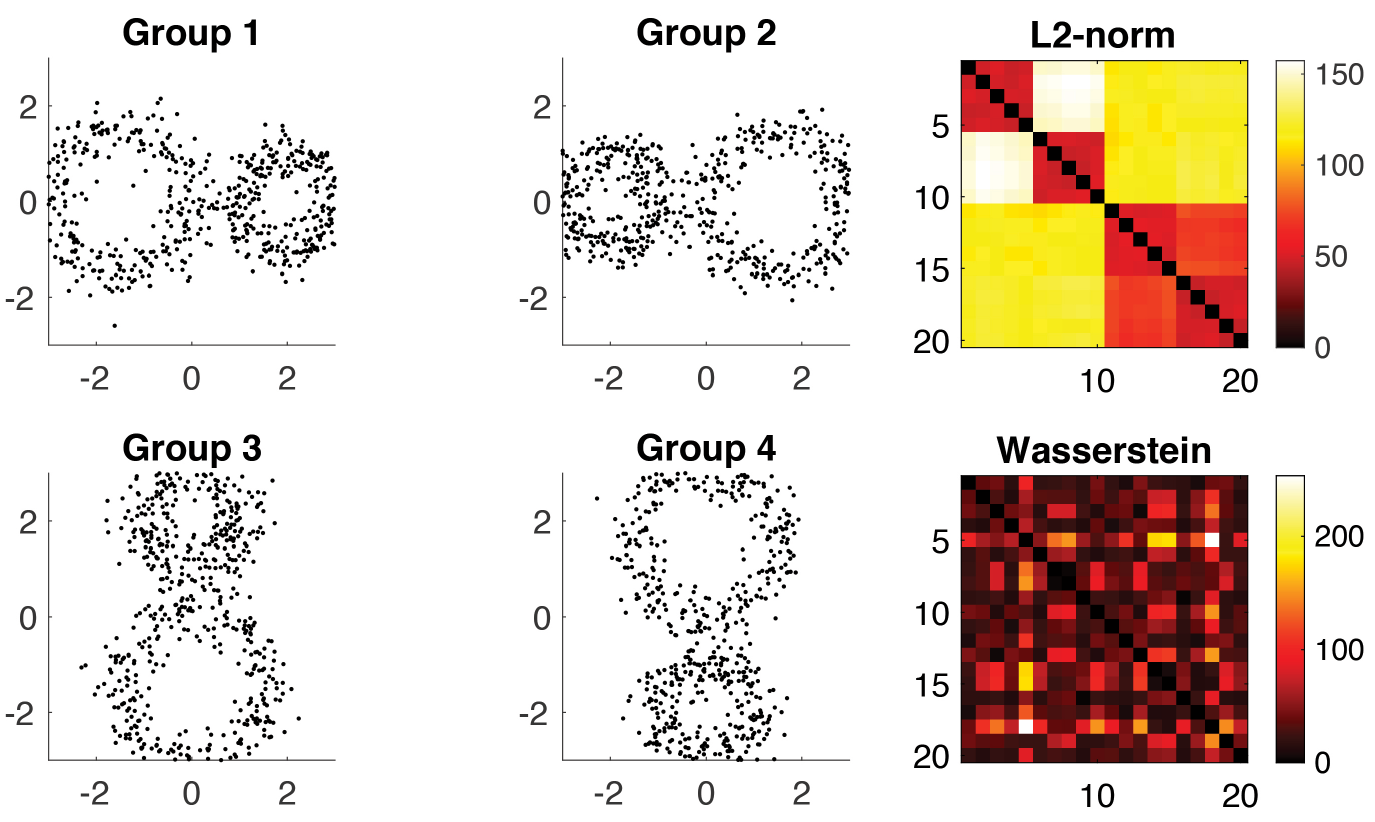}
\caption{
Networks in Groups 2, 3, and 4 are generated by rotating those in Group 1. Since these networks are topologically equivalent, one would not expect to see any clustering pattern in the distance matrix. However, the distance matrix based on the Euclidean distance (L2-norm) exhibits a clustering pattern. In contrast, the topological distance, computed using the Wasserstein distance, does not display any such block pattern.}
\label{fig:simulationnodiff}
\end{figure}

Suppose we have networks \( G_i = (V, w^{i}) \) with a fixed node set $V = \{1, \cdots, q \}$. Let the birth and death sets be
\[
{\bf B}_{G_i}: \epsilon_{b_{1}}^i < \dots < \epsilon_{b_{m_{0}}}^i, \quad {\bf D}_{G_i}: \epsilon_{d_{1}}^i < \dots < \epsilon_{d_{m_{1}}}^i.
\]
Then, the 2-Wasserstein distance for 0D homology (connected components)  is given by
$$
d_0(G_1, G_2) = \sum_{i=1}^{m_0} [\epsilon_{b_i}^1 - \epsilon_{b_i}^2]^{2}.
$$
Similarly, the 2-Wasserstein distance for 1D homology (loops) is  given by
$$
d_1(G_1, G_2) = \sum_{i=1}^{m_1} [\epsilon_{d_i}^1 - \epsilon_{d_i}^2]^{2}.
$$
It is possible to combine 0D and 1D topological distances as
$$d (G_1, G_2) = w_0 d_0(G_1, G_2) + w_1 d_1(G_1, G_2).$$
In this study, we will simply use the equal weights $w_0 = w_1 =1$. 
The 2-Wasserstein distances are computed using  a MATLAB function call {\tt M = WS\_pdist2(C\_1,C\_2)}, which inputs a collection of connectivity matrices {\tt C\_1} of size $q \times q \times m$ and {\tt C\_2} of size $q \times q \times n$. $q$ is the number of nodes and $m$ and $n$ are the samples in two groups. Then the function outputs structured array {\tt dist}, where
{\tt M.D0}, {\tt M.D1} and {\tt M.D01} are $(m+n) \times (m+n)$ pairwise distance matrix for 0D distance $d_0$, 1D distance $d_1$, combined distance $d = d_0 + d_1$ respectively.

To see the effect of the Wasserstein distance, we generated 4 circular patterns of identical topology (Figure \ref{fig:simulationnodiff}). Along the circles, we uniformly sampled 60 nodes and added  Gaussian noise $N(0, 0.3^2)$ on the coordinates. We generated 5 random networks per group. The Euclidean distance ($L_2$-norm)  between randomly generated points are used to build connectivity matrices. Figure \ref{fig:simulationnodiff} displays the superposition of nodes from 5 networks in each group.  Since they are topologically equivalent, the distance between networks should show no clustering pattern. In fact the Wasserstein distance $d=d_0 + d_1$ shows no discernible clustering pattern while $L_2$-norm shows the clustering pattern.  The $L_2$-norm distance is particularly large between horizontal (Groups 1 and 2) and vertical patterns (Groups 3 and 4).

\subsection{Online  topological inference on distance matrix}
\label{sec:inference}

Assume we have two groups of networks \(C_1 = \{X_1, \dots, X_m\}\) and \(C_2 = \{Y_1, \dots, Y_n\}\). If there is a group difference, the topological distances are expected to be relatively small within groups and relatively large between groups. The  topological distance within the groups is given by
\[
l_{W} = \sum_{i,j} d(X_i, X_j) + \sum_{i,j} d(Y_i, Y_j).
\]
Similarly, the topological distance between the groups is given by
\[
l_{B}  = \sum_{i,j} d(X_i, Y_j).
\]
Figure \ref{fig:method1} shows a schematic of between- and within-group distance computation. Although we restrict the inference to a two-sample comparison setting, the inference can be easily generalized to an arbitrary number of groups. We then use the ratio statistic
\[
\phi  = \frac{l_B }{l_{W} }
\]
for testing the topological difference between the groups of networks. If \(\phi\) is large, the groups differ significantly in network topology. If \(\phi\) is small, the group difference is small. Since the distribution of the ratio statistic \(\phi\) is unknown, the permutation test is used to determine the empirical distributions. To speed up the computation, we adapted a scalable online computation strategy through the {\em transposition test} as follows \cite{chung.2019.CNI}.

We first merge two groups and create a distance matrix with dimensions \((m + n) \times (m + n)\), covering all network pairs. Then, we apply a permutation test by shuffling the rows and columns of the distance matrix based on permuted group labels. This avoids the need to recalculate distances and speeds up the process. To further accelerate the computation, we employ the transposition test, an efficient variant of the permutation test \cite{song.2023,chung.2019.CNI}. In this test, we focus on how the within-group \(l_W\) and between-group \(l_B\) distances change when we swap only one entry from each group through a \textit{transposition}. 
Assume we swap the $k$-th and $j$-th entries between the groups. After each transposition, the within-group distance changes as:
\[
l_W' = l_W + \Delta_{W},
\]
where \(\Delta_{W}\) represents the entries that need to be swapped. This requires swapping only \(\mathcal{O}(m+n)\) entries, in contrast to the \(\mathcal{O}((m+n)^2)\) entries needed in a standard permutation test. Similarly, the between-group distance changes as:
\[
l_B'= l_B + \Delta_{B}.
\]
The ratio statistic is then updated sequentially over random transpositions from $\phi = l_B /l_W$ to $\phi' = l_B' / l_W'$. The algebraic details on $\Delta_{W}$ and $\Delta_{B}$ are given in \cite{song.2023}.

\begin{figure}[t!]
	\centering
	\includegraphics[width=1\linewidth]{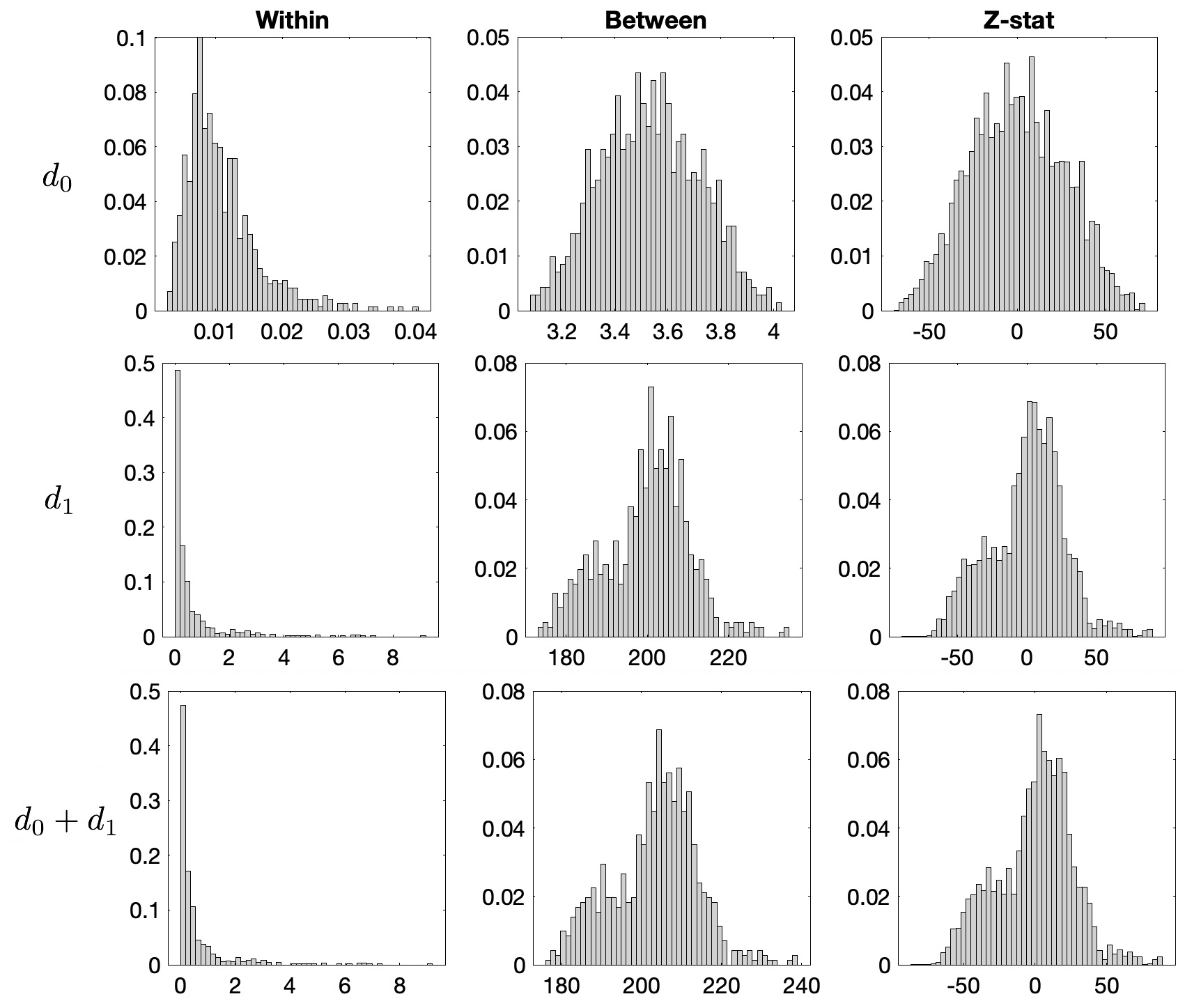}
	\caption{The distribution of within- and between-group distances obtained from Jackknife resampled structural covariance networks. The within- and between-group distances are statistically independent and thus we can compute the Z-statistic out of the distances.}
	\label{fig:dist}
\end{figure}

In numerical implementation, to mitigate potential bias and hasten convergence, we intersperse a full permutation among every 1000 transpositions. Figure \ref{fig:dist} shows distributions of within- and between-group distance and the convergence plot of the transposition test. Our approach does not assume any specific distribution for the test statistic, making it robust against varying variances between groups. Like the standard permutation test, the transposition test approximates the null distribution of the test statistic, allowing us to quantify deviations in the observed data from the null distribution \cite{bullmore.1999, chung.2018.rapid, hayasaka.2004, nichols.2002}.

\subsection{Z-statistic between between- and within-group distances}

We can also develop a $Z$-test like parametric test procedure based on a Gaussian distribution testing difference in the between- and within-group distances. Let $L_W$ be a pairwise within-group distance, which is random, realized by every possible $d(X_i,X_j)$. 
Let $L_B$ be a pairwise between-group distance, which is random, realized by every possible $d(X_i,Y_j)$. Then 
the average pairwise within-group distance is given by
\[ \mathbb{E} L_W =\frac{\sum_{i,j} d (X_i,X_j) + \sum_{i,j}  d(Y_i, Y_j)}{m(m-1) + n(n-1)}. 
 \]
 The second moment of $L_W$ is given by 
\[ \mathbb{E} L_W^2 = \frac{\sum_{i,j} d^2(X_i,X_j) + \sum_{i,j} d^2(Y_i, Y_j)}{m(m-1) + n(n-1)}.
\]
The variance is given by $\mathbb{V} L_W =  \mathbb{E} L_W^2 - (\mathbb{E} L_W)^2$. Similarly, the average pairwise between-group distance is given by
\[ \mathbb{E}  L_{B} =\frac{\sum_{i,j} d(X_i, Y_j)}{mn} .\]
The second moment is given by
\[
\mathbb{E} L_B^2 = \frac{\sum_{i,j} d^2(X_i, Y_j)}{mn}.
\]
The variance is given by $\mathbb{V} L_B =  \mathbb{E} L_B^2 - (\mathbb{E} L_B)^2$.
Assuming two groups $C_1$ and $C_2$ are independent samples, the distances $d (X_i,X_j)$ and $d(Y_e,Y_f)$ are independent. 
The distance $d(X_i,X_j)$ is also {\em conditionally} independent of $d(X_i,Y_f)$ over fixed $X_i$. Since we have the conditional independence for every possible $X_i \in C_1$, $d(X_i,X_j)$ and $d(X_i,Y_f)$ are   
 independent. Following the similar logic,  $d (X_i,X_j)$  and $d(X_e,Y_f)$ are also independent. 

Subsequently, the within- and between group distances are independent. Then the $Z$-statistic of  two independent random variables $L_B$ and and $L_W$  is given by
$$Z = \frac{L_B - L_W - ( \mathbb{E} L_B - \mathbb{E} L_W)}{\sqrt{ \frac{\mathbb{V} L_B}{mn} + \frac{\mathbb{V} L_W}{ m(m-1)+n(n-1)}   }}$$
Then we are testing the null hypothesis
$$H_0:  \mathbb{E} L_B = \mathbb{E} L_W$$
against the alternative
$$H_1:  \mathbb{E} L_B \geq \mathbb{E} L_W.$$

The between-group distance is expected to be larger than the within-group distance. Under the null hypothesis, \( Z \) should asymptotically follow the standard normal distribution \( N(0,1) \).  Figure \ref{fig:dist} displays the distributions of within- and between-group distances for each topological distance used in our study.

\section{Application}

\subsection{Imaging data and pre-processing}

The study included 23 children who suffered maltreatment in early life, and 31 age matched typically developing comparison children \cite{chung2013persistent,chung2015persistent,hanson2013early}. 
All subjects were scanned at the University of Wisconsin-Madison. The maltreated sample suffered early childhood neglect as they were initially raised in institutional setting; in such settings, there is a lack of toys or simulation, unresponsive caregiving, and an overall dearth of individualized care and attention \cite{rutter1998developmental}. These children were, however, then adopted and then move into normative caregiving environments. For the controls, we selected children without a history of maltreatment from families with similar ranges of socioeconomic statuses. The exclusion criteria include, among many others, congenital abnormalities (e.g., Down syndrome or cerebral palsy) and fetal alcohol syndrome (FAS). The average age for maltreated children was 11.26 $\pm$ 1.71 years while that of controls was 11.58 $\pm$ 1.61 years. This particular age range was selected since this development period is characterized by major regressive and progressive brain changes \cite{lenroot2006brain,hanson2013early}. There are 10 boys and 13 girls in the maltreated group and 18 boys and 13 girls in the control group. Groups did not statistically differ on age, pubertal stage, sex, or socio-economic status \cite{hanson2013early}. 
The average amount of time spent in institutional care by children was 2.5 years $\pm$ 1.4 years, with a range from 3 months to 5.4 years. Children were on average 3.2 years old $\pm$ 1.9 months when they were adopted, with a range of 3 months to 7.7 years. T1-weighted MRI  were collected using a 3T General Electric SIGNA scanner (Waukesha, WI) with a quadrature birdcage head coil. DTI were also collected in the same scanner using a cardiac-gated, diffusion-weighted, spin-echo, single-shot, EPI pulse sequence \cite{hanson2013early}. Diffusion tensor encoding was achieved using twelve optimum non-collinear encoding directions with a diffusion weighting of 1114 s/mm$^2$ and a non-DW T2-weighted
reference image. Other imaging parameters were TE = 78.2 ms, 3 averages (NEX: magnitude
averaging), and an image acquisition matrix of 120 $\times$ 120 over a field of view of 240 $\times$ 240 mm$^2$. The acquired voxel size of $2 \times  2 \times 3$ mm was interpolated to 0.9375 mm isotropic dimensions (256 $\times$ 256 in plane image matrix). To minimize field inhomogeneity and image artifacts, high order shimming and field map images were collected using a pair of non-EPI gradient echo images at two echo times: TE1 = 8 ms and TE2 = 11 ms.

 \begin{figure}[t!]
	\centering
	\includegraphics[width=22pc]{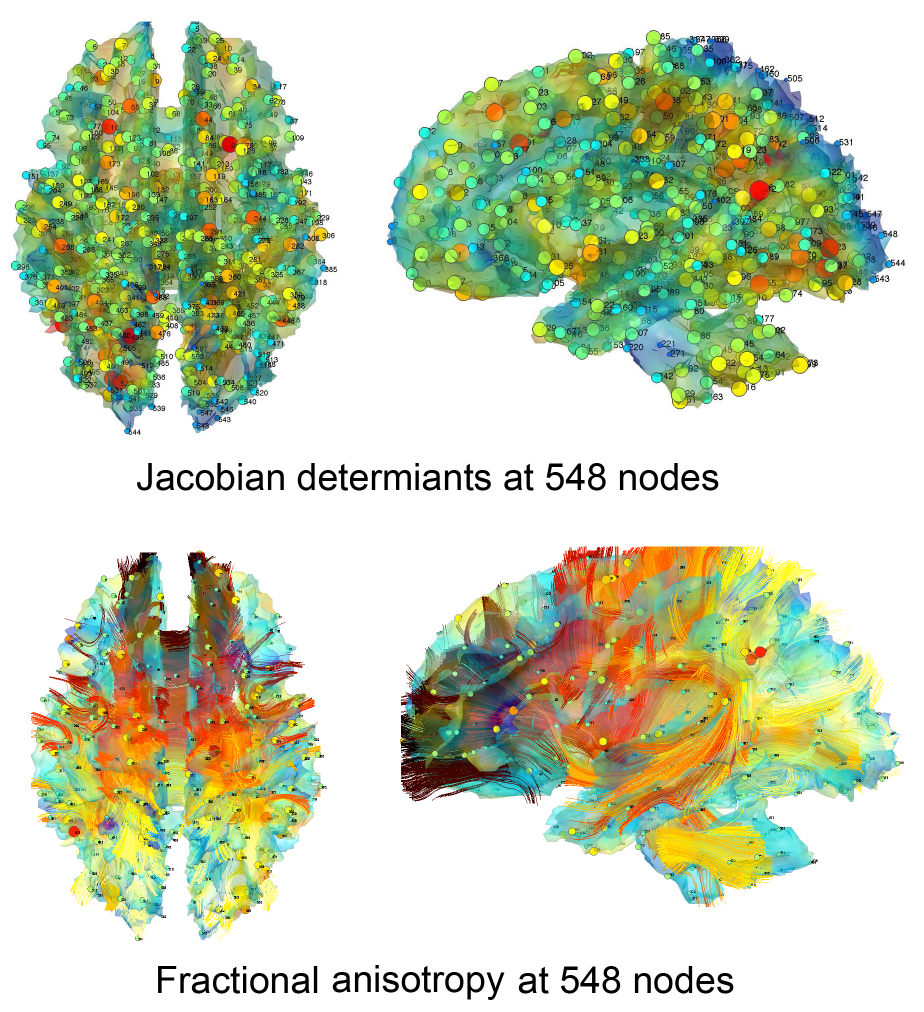}
	\caption{548 uniformly sampled nodes along the white matter surface. The nodes are sparsely sampled on the template white matter surface to guarantee there is no spurious high correlation due to proximity between nodes. The same nodes are taken in both MRI and DTI for comparison between the two modalities. Bottom: curves are extracted white matter fiber tracts from a subject.}
	\label{fig:tstat}
\end{figure}

For T1-MRI, a study specific template was constructed using the diffeomorphic shape and intensity averaging technique through Advanced Normalization Tools (ANTS) \cite{avants2008symmetric}. Image normalization of each individual image to the template was done using symmetric normalization with cross-correlation as the similarity metric. The 1mm resolution inverse deformation fields are then smoothed out with a Gaussian kernel of 4mm (full width at half maximum, FWHM). The Jacobian determinants of the inverse deformations from the template to individual subjects were computed at each voxel. The Jacobian determinants measure  the amount of voxel-wise change from the template to the individual subjects \cite{chung2001unified}. For diffusion-MRI, images were corrected for eddy current related distortion and head motion via FSL software and distortions from field inhomogeneities were corrected using custom software based on the method given in \cite{jezzard1999sources} before performing a non-linear tensor estimation using CAMINO \cite{camino2006open}. Subsequently, we have used iterative tensor  image registration strategy for spatial normalization using DTI-ToolKit \cite{joshi2004unbiased,zhang2007high}. Then fractional anisotropy (FA) values were calculated for diffusion tensor volumes diffeomorphically registered to the study specific template.

White matter was segmented into tissue probability maps using template-based priors and then registered to a study-specific template \cite{bonner2012gray,chung.2015.TMI}. We thresholded the white matter density at a value of 0.7 to obtain an isosurface, which is located within the white matter rather than at the boundary between gray and white matter. Our interest lies in detecting changes along this surface close to the actual tissue boundary. This isosurface was represented as a triangle mesh with 189,536 vertices, resulting in an average inter-nodal distance of 0.98 mm. Given the high correlation between Jacobian determinants and FA values at neighboring voxels, we uniformly sampled the mesh vertices to yield \( q=548 \) nodes, which produced an average inter-nodal distance of 15.7 mm. This distance is sufficiently large to avoid spuriously high correlations between adjacent nodes (see Figure \ref{fig:tstat}). Subsequently, we computed \( 548 \times 548 \) sample correlation matrices across subjects. Functional parcellations such as those by Gordon \cite{gordon.2016} and Schaefer \cite{schaefer.2017} are primarily based on fMRI studies and may not be well-suited for structural covariance networks, which operate at higher spatial resolutions based on anatomical measurements. Furthermore, many existing parcellations focus mainly on gray matter, where DTI measurements such as FA can be difficult to estimate reliably. White matter tracts, reconstructed using tractography algorithms, do not consistently extend all the way to the gray matter, making it challenging to robustly estimate FA values from DTI in these regions \cite{maier.2017}.

\begin{figure}[t!]
\centering
\includegraphics[width=0.8\linewidth]{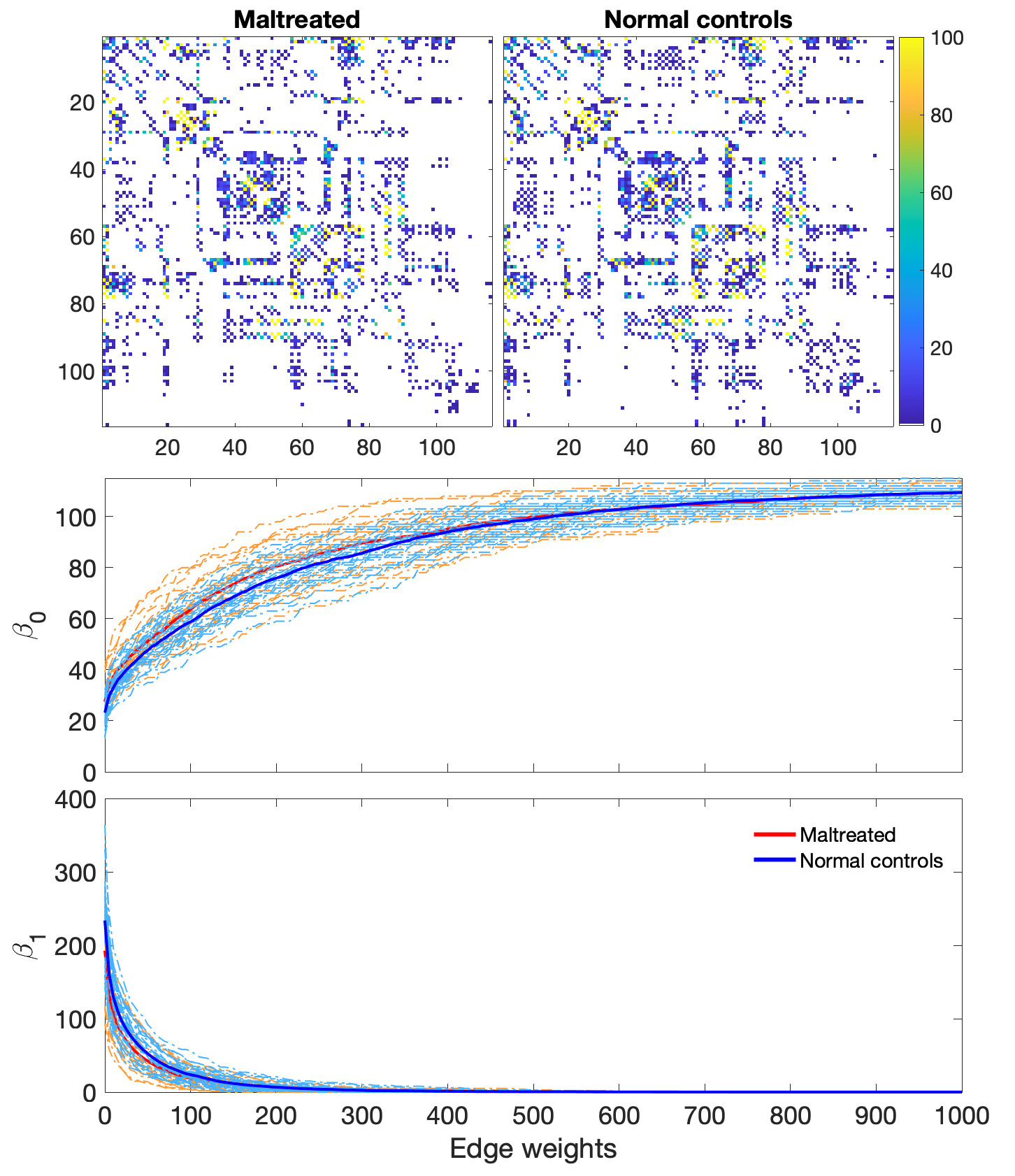}
\caption{Top: The average structural connectivity in maltreated children compared to normal controls. Bottom: Individual Betti curves for each subject are displayed. The thick red and blue curves represent the average Betti curves for the maltreated and control groups, respectively. Given that structural connectivity predominantly forms a single, large connected tree, there is minimal variation in the topological profiles. Thus, no statistically significant topological differences were detected between the groups.}
\label{fig:maltreated-beta}
\end{figure}

\subsection{Structural connectivity analysis}

Tractography was performed in the normalized space using the TEND algorithm and warped into the study template \cite{lazar.2003.hbm}. We utilized the Anatomical Automatic Labeling (AAL) atlas with 116 parcellations \cite{tzourio.2002}. This atlas was registered to the study template via diffeomorphic image registration. The endpoints of fiber tracts were identified with respect to these 116 parcellations, and tracts passing between parcellations were counted. Tracts not passing through two given parcellations were excluded. We applied the proposed topological inference methods to the resulting structural connectivity matrices (Figure \ref{fig:maltreated-beta}). The transposition test was conducted with 1 million transpositions. To accelerate convergence and mitigate potential bias, one permutation was introduced for every sequence of 1000 consecutive transpositions. We did not observe any statistically significant topological differences between the groups. All three topological distances \(d_0\), \(d_1\), and \(d_0 + d_1\) yielded \(p\)-values of 0.56, 0.34, and 0.57, respectively.

Structural connectivity is characterized predominantly by a single, large connected component with few loops \cite{chung.2011.SPIE}. We found that 96\% of all nodes formed a single gigantic connected tree. Thus, structural connectivity is primarily characterized by 0D homology, highlighting the deterministic and hierarchical nature of anatomical pathways between brain regions. Given trees with an identical number of nodes, they are all topologically equivalent. The direct application of TDA methods to structural connectivity matrices, therefore, diminishes statistical power.

For example, consider two different trees \(T_1\) and \(T_2\) with the same \(q\) number of nodes but with sorted, identical edge weights $$w_{(1)} < w_{(2)} < \cdots < w_{(q-1)}.$$ 
When performing graph filtrations on these trees, the resulting 0D and 1D persistence diagrams will be identical. 
The best topological matching between \(T_1\) and \(T_2\) is simply given by matching the $i$-th smallest birth values together. 
Consequently, the 2-Wasserstein distances vanish, i.e., 
$$d_0(T_1, T_2) = d_1(T_1, T_2) = 0,$$
making it impossible to distinguish between the trees.

\subsection{Structural covariance network analysis}

\begin{figure}[t!]
	\centering
	\includegraphics[width=1\linewidth]{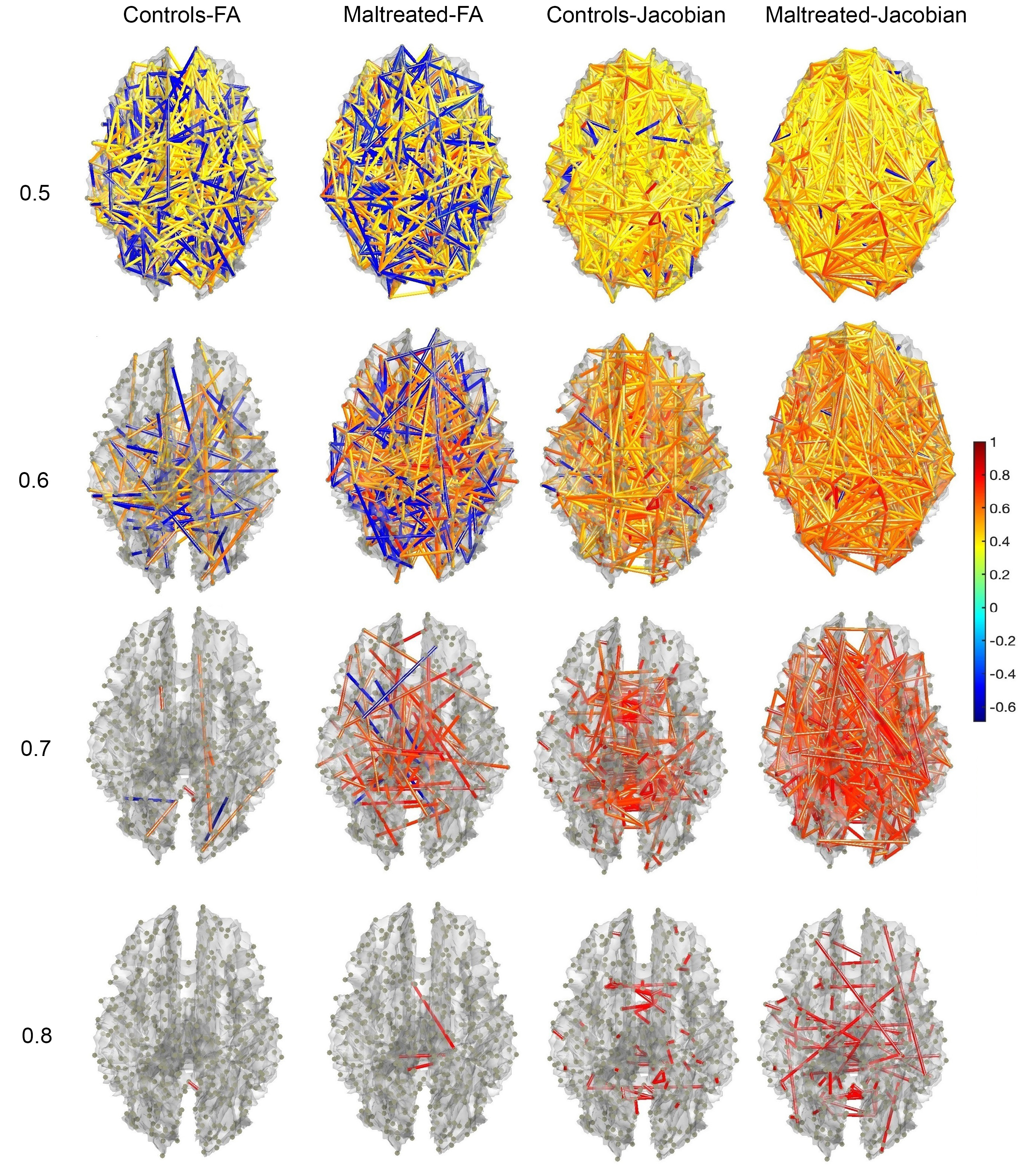}
	\caption{Structural covariance networks on 548 nodes, generated from fractional anisotropy (FA) values derived from DTI and Jacobian determinants derived from T1-MRI. The networks are thresholded at values of 0.5, 0.6, 0.7, and 0.8, shown from top to bottom. The color bar represents the correlation values for each edge.}
	\label{fig:3d}
\end{figure}

We sequentially thresholded the correlation matrices to obtain graph filtrations. Figure \ref{fig:3d} displays the thresholded structural covariance networks at correlation values 0.5, 0.6, 0.7 and 0.8. These networks reveal strongly correlated connections in maltreated children, indicating a highly homogeneous nature of white matter structures in this group. Higher correlation values would be expected if FA and Jacobian determinants are homogeneous within each group.

Since there are only one correlation matrix per group, this gives a challenge in group level topological inference. Thus, 
we adapted the leave-one-out Jackknife resampling to generate multiple correlation matrices per group as follows.
There are $m=$31 normal controls and $n=$23 maltreated children in our sample. For the normal controls, we leave the \(i\)-th subject out and compute the group-level correlation matrix using the remaining 30 subjects, denoting this matrix as \(X_i\). This process is repeated for all subjects to obtain the structural covariance networks \(X_1, \ldots, X_{m}\). Similarly, for the maltreated children, we leave the \(i\)-th subject out and compute the group-level correlation matrix using the remaining 22 subjects, denoting this as \(Y_i\). This process is repeated to obtain \(Y_1, \ldots, Y_n\). These resampled correlation matrices are then feed into the proposed topological data analysis.

\begin{figure}[t!]
	\centering
	\includegraphics[width=1\linewidth]{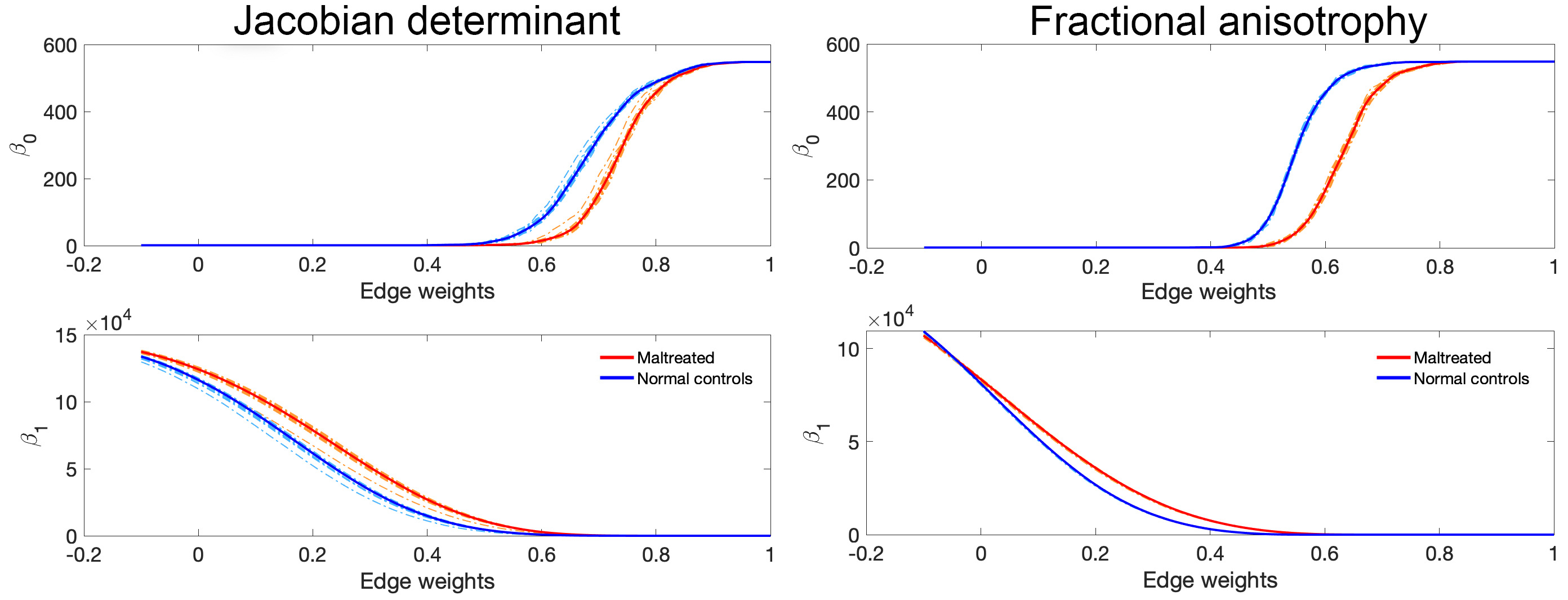}
	\caption{The Betti curves are derived from the Jackknife-resampled structural covariance networks for both the Jacobian determinants (left) and FA-values (right). Compared to the Jacobian determinants, the FA-values exhibit significantly less variability in their topological profiles.
		\label{fig:bettiplot}}
\end{figure}

Using the resampled correlation matrices of the Jacobian determinants and fractional anisotropy (FA) values on 548 nodes, we calculated both the Betti-0 and Betti-1 curves for all subjects (Figure \ref{fig:bettiplot}). For the same filtration values, the Betti-0 curves indicated higher values, i.e., more connected components, in the control group compared to the maltreated group. This observation implies that brain regions in the control group are less correlated across different regions, suggesting a more heterogeneous anatomical structure. This is in contrast to the maltreated group, which exhibited higher Betti-0 curves in the tractography-based connectivity study in the previous section. This suggests a less fractured and more interconnected network in the control group.

On the other hand, the Betti-1 curves for the maltreated group were higher than those for the control group (Figure \ref{fig:bettiplot}). This indicates that maltreated children have more loops, which can only occur if there are denser and more higher correlations in their structural covariance networks. This again points to a more homogeneous nature of the structural covariance networks in maltreated children. The pattern is reversed in the tractography-based connectivity study, where lower Betti-1 curves are observed for the maltreated group. While the loops in the structural covariance networks are statistical in nature, the loops in tractography-based connectivity represent actual physical connections. In summary, by employing Betti-0 and Betti-1 curves, we are able to visualize and characterize the topological differences between the maltreated and control groups, particularly in terms of connected components and loops. These Betti curves may serve as potential biomarkers for distinguishing between maltreated subjects and the control group.

To more rigorously quantify the topological differences, we used the Wasserstein distance based ratio statistic. First, we performed  the Jackknife resampling. Then computed the between-group and within-group Wasserstein distances using $ d_0 $ , $ d_1 $ and $d_0 + d_1$. Figure \ref{fig:dist} displays the distribution of between-group and within-group Wasserstein distances. We notice a significant distinction between the Jackknife resampled Betti curves of both groups which is much larger than within-group variability using all three $ d_0 $, $ d_1 $, and $ d_0 + d_1$ distances and reveals the between group difference is highly significant. Figure \ref{fig:dist} clearly shows that the variability between groups is far larger than within-group variability. The $p$-values are very small ($p$-value $< 0.001$) for $ d_0 $, $ d_1 $, and $ d_0 + d_1$ 
for both Jacobian determinants and FA values. We conclude that  there are significant topological differences in the topological structure of MRI and DTI structural covariance networks. Note our ratio test statistic is global test procedure over the range of filtration values and space so there in no need for multiple comparisons.

We also performed the parametric $Z$-test.  Figure \ref{fig:dist} displays the distributions of within- and between-group distances for the topological distances \( d_0 \), \( d_1 \), and \( d = d_0 + d_1 \) used in our study. The distribution of the \( Z \)-statistic is also displayed for each distance. We evaluated the normality of the \( Z \)-statistic under the null hypothesis using the Kolmogorov-Smirnov (KS) test, which is a non-parametric statistical test used to compare a sample distribution with a reference probability distribution \cite{conover.1980,gibbons.1992}. The statistical significance for all distance metrics were below 0.001, indicating a high likelihood that the \( Z \)-statistic follows a normal distribution. Therefore, we can proceed with parametric tests based on the normal distribution. The resulting \( p \)-values were all below 0.001, indicating statistically significant differences between the groups for all distance metrics.

\section*{Discussion}

To investigate the topological impact of maltreatment on brain networks, we applied TDA methods to structural covariance networks. We observed fewer disconnected components in maltreated children compared to controls (Figure \ref{fig:3d}). This may be attributed to the higher anatomical homogeneity observed in the white matter structure of maltreated children.  \cite{hanson2013early} also noted disrupted white matter organization in neglected children, which resulted in more diffused connections between brain regions. This will likely increase anatomical homogeneity across brain regions. Our topology-based approach successfully revealed these alterations and suggests that TDA could serve as a biomarker for identifying the neurobiological impacts of maltreatment \cite{gomez2009use,jeong2004eeg,besthorn1995parameters,dastgheib2011application}.

The maltreatment and malnutrition often co-occur, typically in the form of neglect. For instance, a caregiver might intentionally or unintentionally fail to provide adequate nutrition, leading to malnutrition and a range of developmental, psychological, and health issues \cite{aber.1984,baer.2006}. Neglect is often the predominant form of maltreatment leading to malnutrition, making malnourished children more susceptible to illness, developmental delays, and in extreme cases, death \cite{perez.1994}. Both maltreatment and malnutrition can have severe and often synergistic neurodevelopmental consequences, affecting regions of the brain responsible for cognitive function and emotional regulation \cite{teicher.2014,teicher.2016}. \cite{teicher.2014} employed structural covariance network analysis using cortical thickness and considered various nodal centrality measures like degree, betweenness, closeness, and eigenvector. The study observed a significant decrease in nodal centralities across most brain regions, except for an increase in the right anterior insular gyrus and right precuneus gyrus. An increase in correlation in structural covariance networks could lead to an increased degree centrality if new edges are formed or existing edges are strengthened. \cite{chung.2017.BC} conducted a study using  DTI to examine the probability distribution of node degrees in maltreated children. The study revealed that maltreated children tend to have a higher concentration of low-degree nodes and fewer hub nodes when compared to controls. This observation is consistent with a potential increase in the Betti-0 number in the DTI connectivity of maltreated children.  This finding contrasts with the higher correlations observed in structural covariance networks in the currently. However, if there is a consistently higher level of correlation leading to homogeneous measurements across all brain regions, such a discrepancy can occur. \cite{puetz.2017} found that maltreated children show significant reductions in global connectivity strength and local connectivity, along with increased path lengths. High correlations in structural covariance networks usually translate into more numerous connections between nodes. This creates more direct routes from one node to another, reducing the need for intermediate steps and thereby shortening the average path length.

Persistent homology offers several strengths for neuroimaging research. PH provides a multi-scale framework that allows for the study of brain networks at various resolutions \cite{lee.2012.tmi}. Unlike traditional approaches that rely on a fixed threshold for connectivity, PH accounts for a range of scales, thereby offering a more comprehensive view of brain topology. PH is sensitive to subtle topological differences between networks, making it particularly useful for identifying early markers of neurological diseases and conditions \cite{chung.2019.NN}. Further, PH does not make strong assumptions about the underlying statistical distribution, making it more robust to noise and artifacts commonly encountered in imaging studies. However, PH is not without its limitations. The computation of persistent homology can be computationally expensive, particularly for large and complex networks \cite{zomorodian.2005}. This computational burden may limit its applicability in real-time or large-scale brain imaging studies. PH can sometimes be too sensitive to small topological features that may not be of clinical relevance. The interpretation of PH features, such as Betti numbers and persistence diagrams, can be challenging without a strong mathematical background, which may limit its widespread adoption in the clinical setting. Future work on PH may focus on optimizing the computational aspects of PH and developing user-friendly software tools to promote its application in clinical research. Integrating PH with other machine learning approaches could further refine our understanding of complex brain networks.

 To develop a clinically accurate diagnostic tool from TDA, we need to extended our study to a larger population size, such as the Adolescent Brain Cognitive Development (ABCD) database, the largest long-term study of brain development and child health in US with more than 100 psychiatric and 11 cognitive measures. In the ABCD database, youth ($n=$11,875) 9-11 years of age were recruited for the study.  This age range is important as it is a period of development critical to an individual's life trajectory. The incidence of psychiatric illnesses, such as attention deficit hyperactivity disorder (ADHD), anxiety, mood disorders, and psychosis, increases through adolescence \cite{paus2008many}. The application of our methods to larger datasets such as the ABCD database is left as a future study.

\section*{Acknowledgements}
This study was supported by  NSF MDS-2010778 and NIH EB022856 and MH133614 to MKC, NIH MH43454 to RJD, NIH MH61285 to SDP. A core grant (P50HD105353) to the Waisman Center from the National Institute of Child Health and Human Development is also acknowledged. We would like to thank Sixtus Dakurah of University of Wisconsin-Madison and Yuan Wang of University of South Carolina for discussion on statistical methods. We also like to thank Vijay Anand of University of Exeter  and Anass El Yaagoubi Bourakna for discussion on validation methods.

\section{Technical Terms}

\textbf{Structural Covariance} refers to the statistical relationship in morphological metrics, such as cortical thickness or volume, between different regions of the brain. This concept, often utilized in neuroimaging studies, was first introduced by Keith Worsley in 2005 \cite{worsley.2005.neural,worsley2005connectivity}. It is instrumental in understanding how different brain areas co-vary in their structural attributes across a population. By examining the extent to which the anatomy of one brain region is related to that of another, structural covariance analysis can reveal patterns of connectivity or co-development.

\textbf{Brith-Death Decomposition} involves  simplifying a weighted graph (the brain network) through graph filtration, where edges are sequentially deleted based on sorted edge weights   \cite{song.2023}. It  reveals how network features like connected components and loops appear (birth) or disappear (death). Births occur when new components emerge. Loops are present from the start and are characterized by their death. The decomposition divides edges into a birth set, which contributes to the formation of new components, and a death set, which completes loops. 

The \textbf{Wasserstein Distance} is a metric that quantifies the dissimilarity between two probability distributions, drawing from the theory of optimal transport. This theory seeks the most efficient way to transform one distribution into another. In the context of persistent homology, the Wasserstein Distance is particularly valuable for measuring topological discrepancies between features across various filtrations. For the graph filtration, its capacity for scalable computations makes it an essential tool in the analysis of complex data structures \cite{chung.2023.NI}. \\

The \textbf{Jacobian Determinant} is a key metric in tensor-based morphometry (TBM) for analyzing local volume changes in brain structures. In TBM, brain images from different individuals are nonlinearly mapped onto a common template to identify anatomical variations. The Jacobian determinant is calculated at each voxel of the transformed image, reflecting the local volumetric change at that voxel in comparison to the template. A Jacobian determinant value greater than one signifies local expansion, whereas a value less than one indicates local contraction \cite{chung.2001.NI}.

\bibliographystyle{abbrv}


\bibliography{reference.2022.11.20,reference.2023.11.14}

\end{document}